\documentclass[review,sort&compress,12pt]{elsarticle}

\usepackage{lineno,hyperref}
\usepackage{graphicx}
\usepackage{booktabs}
\usepackage{amssymb}
\usepackage{amsmath}
\usepackage{color,soul}
\usepackage{subfigure}

\begin{document}
	
	\begin{frontmatter}
	%\let\printorcid\relax
	%\let\WriteBookmarks\relax
	%\def\floatpagepagefraction{1}
	%\def\textpagefraction{.001}

	%\shortauthors{E.S. Simakov, A.V. Tyukhtin}  
	
	%\ExplSyntaxOn
	%\keys_set:nn { stm / mktitle } { nologo }
	%\ExplSyntaxOff
	
	%\title [mode = title]{Radiation of a charged particle bunch moving along a deeply corrugated structure with a relatively small period}  

	%\author{E.S. Simakov}
	%\cormark[1]
	%\ead{melo15002009@yandex.ru}

	%\address{Saint Petersburg State University, 7/9 Universitetskaya nab., St. Petersburg 199034, Russia}
	
	\title{Radiation of a charged particle bunch \\ passing through a deeply corrugated structure \\ with a relatively small period}
	
	\author[spbu]{Evgenii S. Simakov\corref{mycorrespondingauthor}}
	\cortext[mycorrespondingauthor]{Corresponding author}
	\ead{st040497@student.spbu.ru}
	
	\author[spbu]{Andrey V. Tyukhtin}
	
	\address[spbu]{Saint Petersburg State University, 7/9 Universitetskaya nab., St. Petersburg, 199034 Russia}
	
	\begin{abstract}
		We analyze electromagnetic radiation from a bunch passing through a corrugated conductive planar structure. 
		The wavelengths of the radiation are assumed to be much greater than the corrugation period. 
		Under this approximation, the corrugated structure can be replaced with a smooth surface on which the so-called equivalent boundary conditions (EBC) are fulfilled. 
		Here, we also assume that the structure is deeply corrugated, i.e. the structure depth is of the same order as an inverse wavenumber. 
		Using the EBC we obtain a general solution and investigate it asymptotically. 
		It is shown that two types of radiation are generated: volume radiation and surface waves.
		Both types of the radiations are analyzed in detail. 
		We demonstrate that the radiation is highly sensitive to the structure depth, which can be used, specifically, for generating the powerful surface radiation. 
		The energy losses of the bunch by both volume and surface radiations are considered as well.
		In particular, the dependences of the energy density on the corrugation parameters are obtained and analyzed.
	\end{abstract}
	
\end{frontmatter}
	
	
%	\linenumbers
	
	\section{Introduction}
	
	The present work aims to study electromagnetic radiation from a charged particle bunch moving in the presence of a corrugated conductive structure.
	In contrast to the traditional consideration when the wavelengths are comparable to a structure period (at these frequencies the well-known Smith-Purcell radiation is generated~%
	\cite{Pot1,Pot2}), here we analyze the so-called ``longwave'' range when the wavelengths are much greater than the period of corrugation.
	This part of the spectrum is not yet well-understood and takes an increasing interest in the scientific community.
	One can note only few relevant studies concerning the analysis of the ``longwave'' range (see, for example,~%
	\cite{BaneStup1,BaneStup2,GinzMalk1,AVVV1,AVVV2,AVAkh,ESAV1,ESAV2,ESAV3}).
	
	For some structures, there are approaches which allow obtaining an analytical solution of a ``longwave'' problem.
	In particular, the method of equivalent boundary conditions (EBC) can be used~%
	\cite{NefSiv}. 
	In short, this method allows replacing a corrugated structure with some anisotropic surface which is characterized by a certain matrix impedance.	
	Recently, we have solved a series of the ``longwave'' problems where the EBC method was applied~%
	\cite{AVAkh,ESAV1,ESAV2,ESAV3}.
	Papers~%
	\cite{AVAkh,ESAV1,ESAV2} are devoted to studying the case of shallow corrugated structures, i.e. when the wavelengths are much greater than the structure depth.
	In~%
	\cite{AVAkh}, we studied the radiation from a bunch moving in a shallow corrugated waveguide.
	Particularly, this work contains a comparison of the theory with simulations carried out in the CST Particle Studio.
	The results of the comparison show that the EBC can be effectively applied even when the wavelengths are only several times greater than a structure period.
	In~%
	\cite{ESAV1,ESAV2}, we analyzed the influence of a shallow corrugated planar structure on the electromagnetic field of a bunch and examined two different cases of a bunch motion (along the structure and perpendicularly to it).
	It was demonstrated that in both cases the bunch generates surface waves propagating in the plane of the structure.
	In particular, we showed that, under certain conditions, the configuration of the surface wave can be used for the determination of the bunch size.
	In contrast to~%
	\cite{AVAkh,ESAV1,ESAV2}, paper~%
	\cite{ESAV3} focuses on studying the case of deep corrugation when the structure depth is much greater than its period (more precisely, it is assumed that the depth is of the same order as an inverse wavenumber).
	In~%
	\cite{ESAV3}, we investigated a situation when a bunch moves along a deeply corrugated planar structure.
	It was shown that the bunch generates surface radiation, and the configuration of the surface wave is rather more complicated than in~%
	\cite{ESAV1,ESAV2}.
	The possibility of applying deeply corrugated structures to bunch diagnostics was discussed as well.	
	
	In the present paper, the structure under consideration is the same as in~%
	\cite{ESAV3}, i.e. it is the deeply corrugated planar surface.
	However, in distinct to~%
	\cite{ESAV3}, here we examine the situation when the bunch passes through the structure (perpendicularly to its plane).
	The general solution of the problem is obtained by the use of the EBC method.
	Then, we carry out the asymptotic analysis of this solution and obtain the components of the electromagnetic field in the form of Fourier-integrals.
	Finally, we consider the energy losses of the bunch by the radiation and analyze the dependences of the energy losses on the problem parameters.

	\section{Equivalent boundary conditions}
	
\begin{figure}[h]
		\centering
		\includegraphics[]{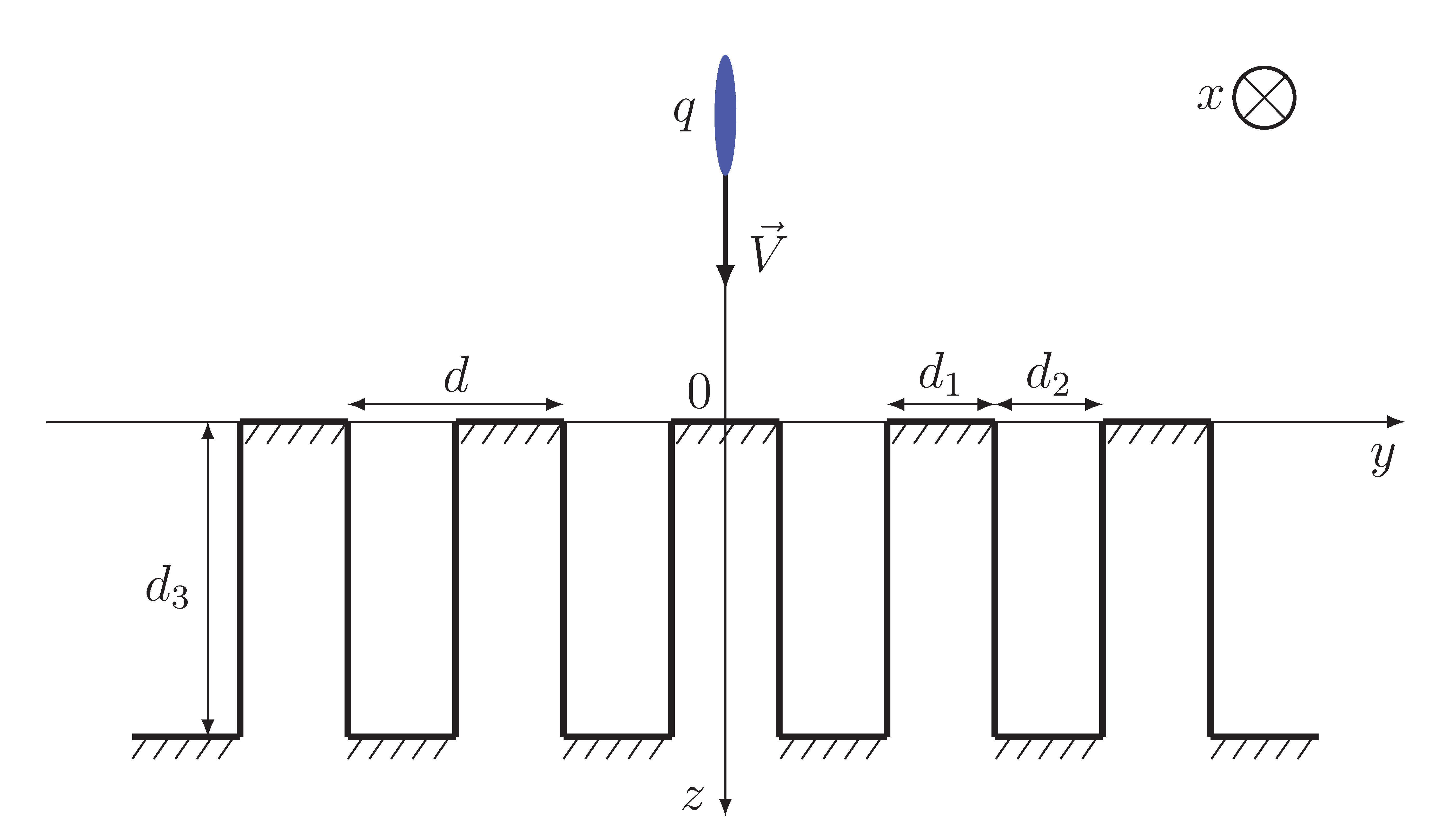}
		\caption[]{The corrugated surface and a moving charge.}
		\label{Fig:1} 
	\end{figure}

	The structure under consideration is a perfectly conductive planar surface having rectangular corrugation (Fig.~%
	\ref{Fig:1}).
	The surface is in a vacuum.
	We study the case when structure period $d$ is much less than emission \linebreak wavelength $\lambda$: 
	\begin{equation}
		\label{eq:2.1}
		\tag{2.1}
		d\ll\lambda.
	\end{equation}
	In this situation, the corrugated structure can be replaced with a smooth surface on which the so-called equivalent boundary conditions (EBC) are fulfilled~%
	\cite{NefSiv}.
	In fact, we deal with an anisotropic surface characterized by certain matrix impedance.

\begin{figure}[h]
		\centering
		\includegraphics[]{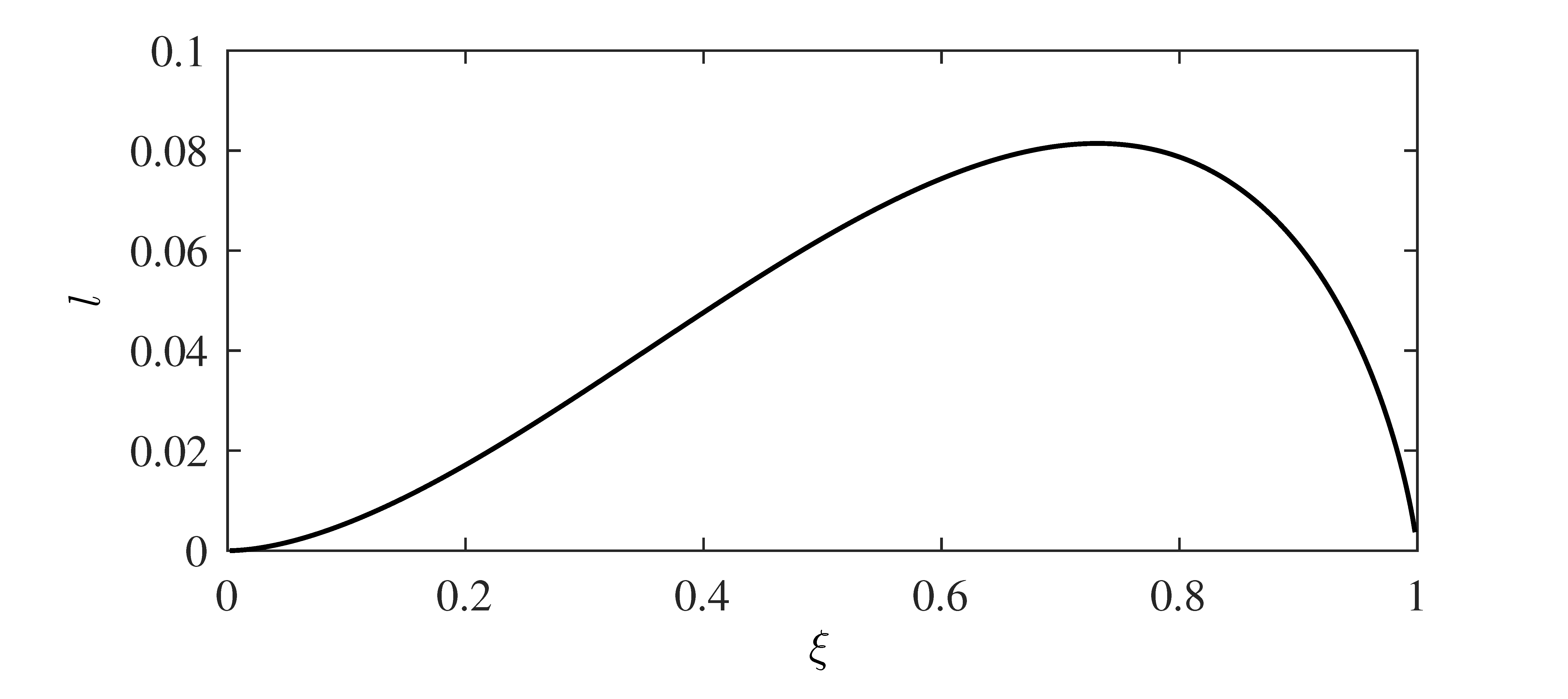}
		\vspace{-15pt}
		\caption{\label{Fig:2} The dependence $l\left(\xi\right)$.} 
	\end{figure}

	Here, we study the case of deep corrugation, i.e. it is assumed that
	\begin{equation}
		\label{eq:2.2}
		\tag{2.2}
		k_0d_3\sim 1,
	\end{equation}
	where $d_3$ is a corrugation depth, $k_0=\omega/c=2\pi/\lambda$ is a wavenumber.
	In this approximation, the EBC have the following form for the Fourier-transforms of electric and magnetic fields on the plane $z=-0$~%
	\cite{NefSiv}:
	\begin{equation}
		\label{eq:2.3}
		\tag{2.3}
		E_{y\omega}\big|_{y=0}=\eta^m H_{x\omega}\big|_{y=0}, \; E_{x\omega}\big|_{y=0}=0.
	\end{equation}
	In~%
	\eqref{eq:2.3}, $\eta^m$ is an impedance which is given by the expression
	\begin{equation}
		\label{eq:2.4}
		\tag{2.4}
		\eta^m=i\frac{d_2}{d}\frac{\operatorname{tg}\left(k_0d_3\right)}{1-k_0d\;l\operatorname{tg}\left(k_0d_3\right)},
	\end{equation}
	where $d_2$ is the width of the structure grooves.
	Parameter $l$ is determined by the formula~%
	\cite{NefSiv}
	\begin{align}
		\label{eq:2.5}
		\tag{2.5}
		l=\frac{1}{2\pi}\left[\left(2-\xi\right)^2\operatorname{ln}\left(2-\xi\right)-\xi^2\operatorname{ln}\xi-2\left(1-\xi\right)\operatorname{ln}4\left(1-\xi\right)\right],
	\end{align}
	where $\xi=d_1/d$, $d_1=d-d_2$.
	Dependence $l\left(\xi\right)$ is shown in Fig.~%
	\ref{Fig:2}.
	Note that parameter $l$ is positive and quite small: $0<l\lesssim 0.082$. 

	\section{General solution}
	
	We consider the electromagnetic field from a charged particle bunch $q$ passing through the corrugated surface with the constant velocity $\mathbf{V}=c\beta\mathbf{e}_z$ which is perpendicular to the structure plane (Fig.~%
	\ref{Fig:1}).
	The bunch is assumed to be infinitely thin in the plane transverse to the direction of the motion and has some longitudinal distribution $\kappa(z-Vt)$, i.e. charge density $\rho$ and current density $\mathbf{j}$ can be written in the form
	\begin{equation}
		\label{eq:3.1}
		\tag{3.1}
		\rho=q\delta(x)\delta(y)\kappa(z-Vt),\;\;\mathbf{j}=\rho\mathbf{V},
	\end{equation}
	where $\delta(\xi)$ is Dirac delta function.
	
	We use Hertz potential $\mathbf{\Pi}$ and present it as a sum of ``forced'' potential $\mathbf{\Pi}^{(i)}$ and ``free'' potential $\mathbf{\Pi}^{(r)}$:
	\begin{equation}
		\label{eq:3.2}
		\tag{3.2}
		\mathbf{\Pi}=\mathbf{\Pi}^{(i)}+\mathbf{\Pi}^{(r)}=\int\limits_{-\infty}^{+\infty}\left(\mathbf{\Pi}_\omega^{(i)}+\mathbf{\Pi}_\omega^{(r)}\right)\exp\left(-i\omega t\right)d\omega,
	\end{equation}
	where $\mathbf{\Pi}_\omega^{(i)}$ and $\mathbf{\Pi}_\omega^{(r)}$ are corresponding Fourier-transforms. 
	The ``forced'' potential describes the electromagnetic field of the bunch in an unbounded vacuum, whereas the ``free'' potential describes an additional electromagnetic field caused by the influence of the corrugated structure (the terms ``forced'' and ``free'' were used by V.L. Ginzburg in~%
	\cite{GinzTsyt1}). 
	Relations between the Fourier-transforms of an electromagnetic field and a Hertz potential are given by the formulas
	\begin{equation}
		\label{eq:3.3}
		\tag{3.3}
		\mathbf{E}_\omega=\mathbf{\nabla}\operatorname{div}\mathbf{\Pi}_\omega+k_0^2\mathbf{\Pi}_\omega, \; \mathbf{H}_\omega=-ik_0\operatorname{rot}\mathbf{\Pi}_\omega.
	\end{equation}

	As follows from Maxwell's equations, the equation for the Hertz potential is
	\begin{equation}
		\label{eq:3.4}
		\tag{3.4}
		\left(\Delta+k_0^2\right)\mathbf{\Pi}_\omega=-\frac{4\pi i}{ck_0}\mathbf{j}_\omega,
	\end{equation}
	where $\mathbf{j}_\omega$ is the Fourier-transform of current density $\mathbf{j}$.
	The solution of Eq.~%
	\eqref{eq:3.4} for the ``forced'' field is the well-known Coulomb field of a moving charge in an unbounded vacuum.
	The Hertz potential of this field is the following:
	\begin{align}
		\label{eq:3.5}
		&\Pi_{x\omega}^{(i)}=\Pi_{y\omega}^{(i)}=0, \nonumber \\ 
		\tag{3.5}
		&\Pi_{z\omega}^{(i)}=\frac{iq\tilde\kappa}{\pi ck_0}\exp\left(i\frac{k_0z}{\beta}\right)\iint\limits_{-\infty \;\;\;}^{\;\;\;\;+\infty}dk_ydk_x\frac{\exp\left(ik_xx+ik_yy\right)}{k_x^2+k_y^2+k_0^2\frac{1-\beta^2}{\beta^2}},
	\end{align}
	where 
	\begin{equation}
		\label{eq:3.6}
		\tag{3.6}
		\tilde\kappa=\frac{1}{2\pi}\int\limits_{-\infty}^{+\infty}d\zeta\kappa\left(\zeta\right)\exp\left(-i\frac{k_0}{\beta}\zeta\right)
	\end{equation}
	is the Fourier-transform of longitudinal charge distribution $\kappa(\zeta)$, $\zeta=z-Vt$.
	
	The ``free'' field can be described by the use of the Hertz potential with two non-zero components:
	\begin{equation}
		\label{eq:3.7}
		\tag{3.7}
		\mathbf{\Pi}_\omega^{(r)}=\Pi_{x\omega}^{(r)}\mathbf{e}_x+\Pi_{z\omega}^{(r)}\mathbf{e}_z.
	\end{equation}
	Potential~%
	\eqref{eq:3.7} should be written as an expansion of waves outgoing from the boundary $z=0$.
	This results in the following expressions for the components:
	\begin{align}
		\label{eq:3.8}
		\tag{3.8}
		\left\{\begin{aligned}
			\Pi_{x\omega}^{(r)} \atop
			\Pi_{z\omega}^{(r)}
		\end{aligned}\right\}{=}\frac{iq\tilde\kappa}{\pi ck_0^2}\!\iint\limits_{-\infty \;\;\;}^{\;\;\;\;+\infty}dk_ydk_x\left\{\begin{aligned}R_x \atop
	    R_z\end{aligned}\right\}\frac{\exp\left[ik_xx{+}ik_yy+ik_{z0}|z|\right]}{k_{z0}},
	\end{align}
    where $k_{z0}=\sqrt{k_0^2-k_x^2-k_y^2}$ ($\operatorname{sgn}\operatorname{Re}k_{z0}=\operatorname{sgn}k_0=\operatorname{sgn}\omega$ for a positive radicand and $\operatorname{Im}k_{z0}>0$ for a negative one).
	In~%
	\eqref{eq:3.8}, $R_x$ and $R_z$ are unknown coefficients which should be obtained from boundary conditions~%
	\eqref{eq:2.3}.
	Taking into account~%
	\eqref{eq:3.3} and solving the system of corresponding algebraic equations one can obtain the following expressions for the coefficients:
	\begin{align}
		\label{eq:3.9}
		\tag{3.9}
		&R_x=-\frac{i\beta k_0k_xk_{z0}\left(k_0+\beta k_{z0}\right)\eta_0^m}{\left(k_0^2-\beta^2k_{z0}^2\right)\left[k_0k_{z0}-i\left(k_0^2-k_x^2\right)\eta_0^m\right]}, \\
		\label{eq:3.10}
		\tag{3.10}
		&R_z=\frac{\beta k_0k_{z0}\left[k_0^2+i\beta\left(k_0^2-k_x^2\right)\eta_0^m\right]}{\left(k_0^2-\beta^2k_{z0}^2\right)\left[k_0k_{z0}-i\left(k_0^2-k_x^2\right)\eta_0^m\right]},
	\end{align}
	where $\eta^m_0=\operatorname{Im}\eta^m$.
	
	Further, we perform the asymptotic analysis of the ``free'' field.
	The asymptotic behavior of Fourier-integrals~%
	\eqref{eq:3.8} is investigated by means of the saddle point method, which is the well-known technique (see, for example, monographs~%
	\cite{FelMarc,BleiHand}).
	One can show that under the condition $k_0R\gg1$, where $R=\sqrt{x^2+y^2+z^2}$, the ``free'' field is practically determined by the contributions of the saddle point and poles.
	As we will see, the contribution of the saddle point is volume radiation, and the contributions of the poles gives surface waves.
	
	We also point out that, from now on, only positive frequencies will be considered ($\omega>0$).
	For negative frequencies all formulas are easy to obtain by the rule $F_{-\omega}=F_{\omega}^{^*}$, where the asterisk means complex conjugation.
	This rule is true for the Fourier-transform of any real function $F\left(t\right)$.

	\section{Volume radiation}
	\label{Sec4}
	
	In this section, we consider the volume part of the radiation.
	This field can be calculated by the use of the saddle point method for multiple integrals~%
	\cite{FelMarc}.
	Assuming $k_0R\gg1$ we obtain the following asymptotic estimation of~%
	\eqref{eq:3.8} (the contribution of the saddle point, which will be marked with index ``$v$''):
	\begin{align}
		\label{eq:4.1}
		\tag{4.1}
		\left\{\begin{aligned}
			\Pi_{x\omega}^{(v)} \atop
			\Pi_{z\omega}^{(v)}
		\end{aligned}\right\}{=}\frac{2q\tilde\kappa}{ck_0^2}\left\{\begin{aligned}R_x \atop
			R_z\end{aligned}\right\}\frac{\exp\left(ik_0R\right)}{R}.
	\end{align}
	Using relations~%
	\eqref{eq:3.3} we find the components of the electromagnetic field.
	In the spherical coordinate system
	\begin{equation}
	    \label{eq:4.2}
	    \tag{4.2}
	    x=R\operatorname{sin}\theta\operatorname{cos}\varphi,\; y=R\operatorname{sin}\theta\operatorname{sin}\varphi,\; z=R\operatorname{cos}\theta
    \end{equation}
    they have the form
	\begin{align}
		\notag
		&E_{R\;\omega}^{(v)}=H_{R\;\omega}^{(v)}=0, \\
		\label{eq:4.3}
		\tag{4.3}
		&\!\!\!\left\{\begin{aligned}
			E_{\theta\;\omega}^{(v)} \\
			E_{\varphi\;\omega}^{(v)}\end{aligned}\right\}{=}
		\left\{\begin{gathered}
			H_{\varphi\;\omega}^{(v)} \\
			-H_{\theta\;\omega}^{(v)}\end{gathered}\right\}
		{\simeq} \frac{2q\tilde{\kappa}}{c}
		\frac{\beta|\!\operatorname{cos}\theta|\!\operatorname{sin}\!\theta}{\left(1{-}\beta^2\!\operatorname{cos}^2\!\theta\right)\!f_0^{(v)}}\left\{\begin{gathered}
			f_1^{(v)} \\
			f_2^{(v)}\end{gathered}\right\}\!\frac{\exp\left(ik_0R\right)}{R},
	\end{align}
	where
	\begin{align}
		\label{eq:4.4}
		\tag{4.4}
		&f_0^{(v)}=|\operatorname{cos}\theta|-i\left(1-\operatorname{sin}^2\theta\operatorname{cos}^2\varphi\right)\eta_0^m, \\
		\label{eq:4.5}
		\tag{4.5}
		&f_1^{(v)}=-1-i\eta_0^m\left[\beta\left(1-\operatorname{sin}^2\theta\operatorname{cos}^2\varphi\right)+\operatorname{cos}\theta\operatorname{cos}^2\varphi\left(1+\beta|\operatorname{cos}\theta|\right)\right], \\
		\label{eq:4.6}
		\tag{4.6}
		&f_2^{(v)}=i\operatorname{sin}\varphi\operatorname{cos}\varphi\left(1+\beta|\operatorname{cos}\theta|\right)\eta_0^m.
	\end{align}
	As follows from~%
	\eqref{eq:4.3}, the volume radiation is separated into two polarizations $\left\{E_{\theta}^{(v)},H_{\varphi}^{(v)}\right\}$ and $\left\{E_{\varphi}^{(v)},H_{\theta}^{(v)}\right\}$, whereas the radial components of both electric and magnetic fields are equal to zero.
	
	Below, we analyze the energy losses of the bunch by the volume radiation.
	We calculate an energy flow through a half-sphere of the radius $R=R_0$ over the entire time.
	This results in the following expression for the energy losses:
	\begin{equation}
		\label{eq:4.7}
		\tag{4.7}
		W^{(v)}=R_0^2\int\limits_{-\infty}^{+\infty}dt\int\limits_{0}^{2\pi}d\varphi\int\limits_{\pi/2}^{\pi}d\theta\operatorname{sin}\theta S_R^{(v)},
	\end{equation}
	where
	\begin{equation}
		\label{eq:4.8}
		\tag{4.8}
		S_R^{(v)}=\frac{c}{4\pi}\left(E_{\theta}^{(v)}H_{\varphi}^{(v)}-H_{\theta}^{(v)}E_{\varphi}^{(v)}\right)=\frac{c}{4\pi}\left[\left(E_{\theta}^{(v)}\right)^2+\left(E_{\varphi}^{(v)}\right)^2\right]
	\end{equation}
	is the radial component of Poynting vector.
	Note that angle $\theta$ in~%
	\eqref{eq:4.7} varies between $\pi/2$ and $\pi$ since we consider the half-space $z<0$ only.
	Then, we write the components of the electromagnetic field as the Fourier-integrals $F\left(\vec{r},t\right)=\int_{-\infty}^{+\infty}d\omega F_\omega \exp\left(-i\omega t\right)$, substitute them into~%
	\eqref{eq:4.7},~%
	\eqref{eq:4.8} and, after some transformations, obtain
	\begin{equation}
		\label{eq:4.9}
		\tag{4.9}
		W^{(v)}=\int\limits_{0}^{+\infty}d\omega\int\limits_{0}^{2\pi}d\varphi\int\limits_{\pi/2}^{\pi}d\theta\operatorname{sin}\theta W_{\omega\Omega}^{(v)},
	\end{equation}
	where
	\begin{equation}
		\label{eq:4.10}
		\tag{4.10}
		W_{\omega\Omega}^{(v)}=cR_0^2\left(|E_{\theta\;\omega}^{(v)}|^2+|E_{\varphi\;\omega}^{(v)}|^2\right)
	\end{equation}
	is a spectral angular density of the energy losses by the volume radiation.
	Note that $W_{\omega\Omega}^{(v)}$ is a sum of the spectral angular densities for both polarizations~%
	\eqref{eq:4.7}.
	\begin{figure*}[!h]
		\vspace{-101pt}
		\begin{center}
			\begin{minipage}[h]{0.51\linewidth}
				\includegraphics[]{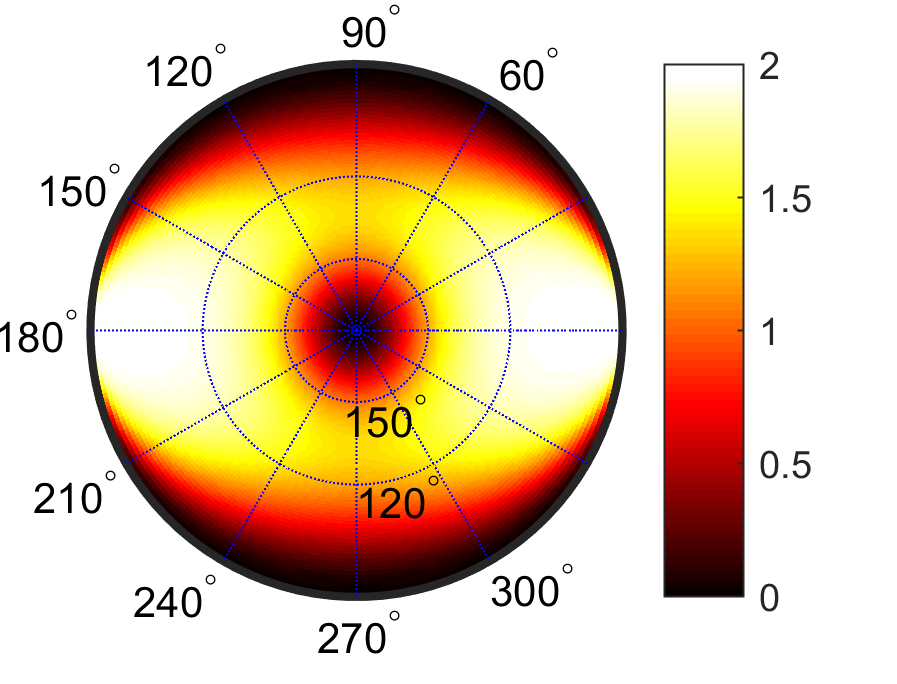}
			\end{minipage}
			\begin{minipage}[h]{0.48\linewidth}
				\includegraphics[]{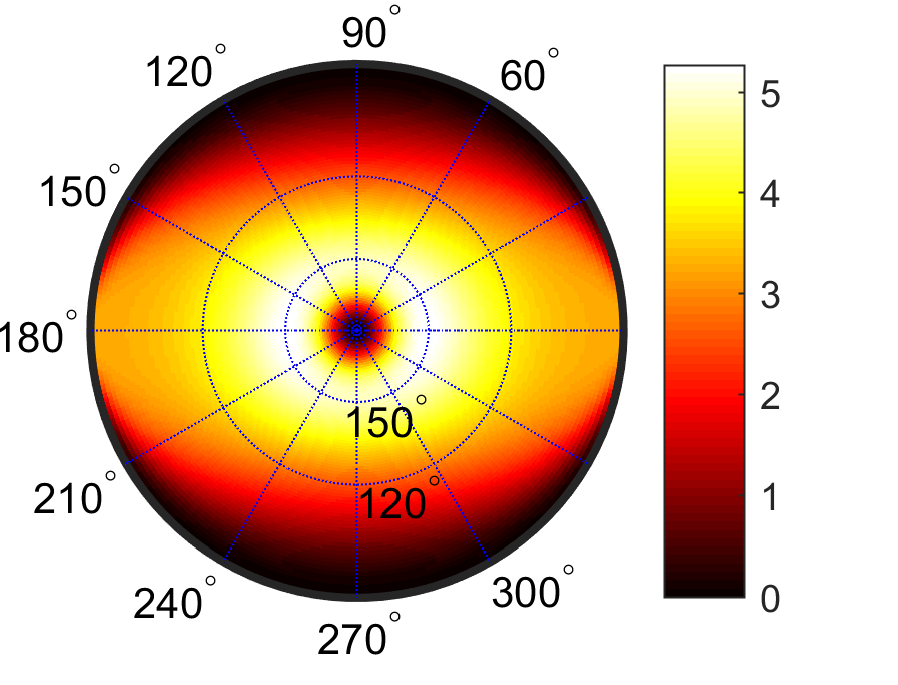}
			\end{minipage}
		    \begin{minipage}[h]{0.51\linewidth}
		    	\includegraphics[]{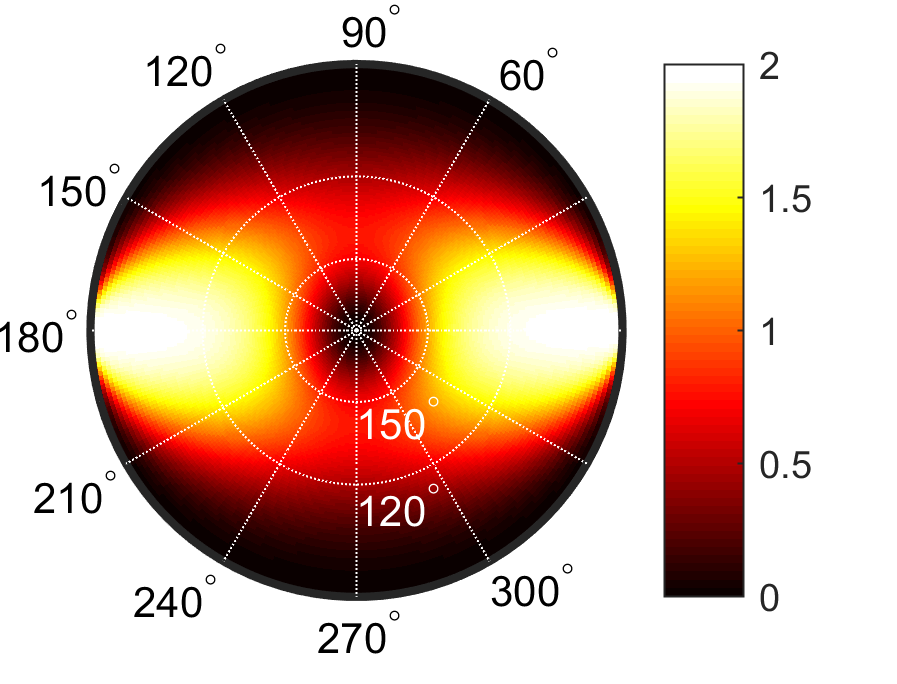}
		    \end{minipage}
		    \begin{minipage}[h]{0.48\linewidth}
		    	\includegraphics[]{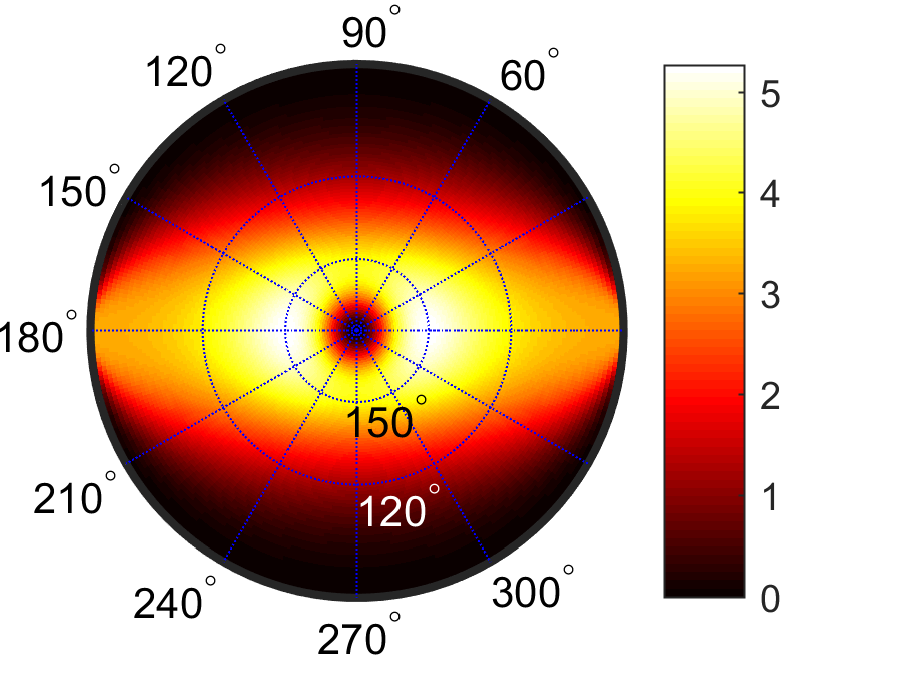}
		    \end{minipage}
	        \begin{minipage}[h]{0.51\linewidth}
	        	\includegraphics[]{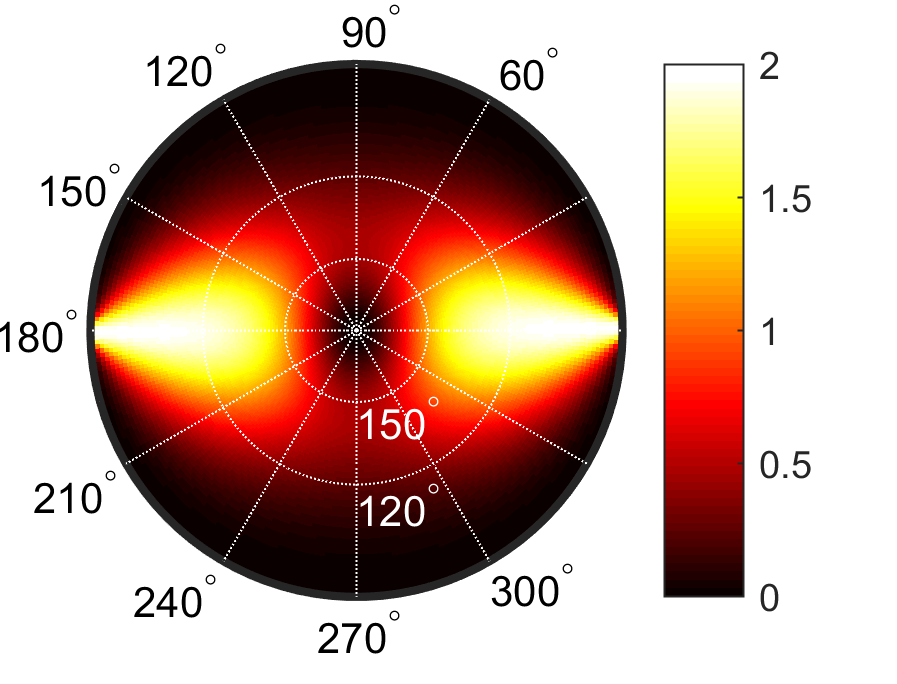}
	        \end{minipage}
	        \begin{minipage}[h]{0.48\linewidth}
	        	\includegraphics[]{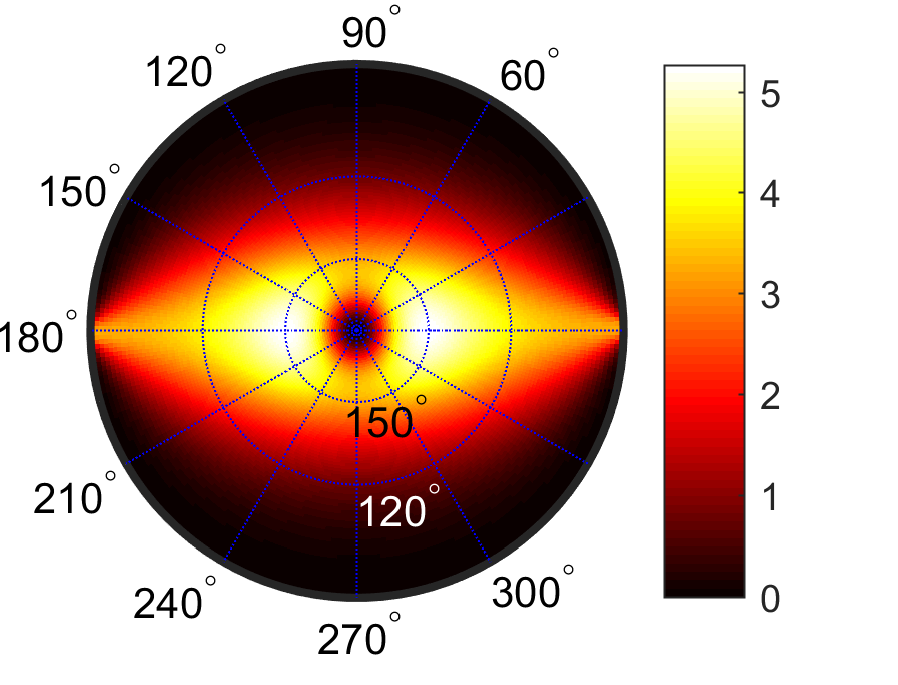}
	        \end{minipage}
		\end{center}
    	\vspace{-15pt}
		\caption{\label{Fig:3} The spectral angular density of the energy losses $W_{\omega\Omega}^{(v)}$ (in $q^2|\tilde{\kappa}|^2/c$ units) depending on the angles of spherical coordinate system~%
		\eqref{eq:4.2}.
		Angle $\theta$ varies along the radius of the circle (from $\theta=90^{\circ}$ at the edge to $\theta=180^{\circ}$ at the center), $\varphi$ is the polar angle.
	    The bunch velocity is $\beta=0.7$ (the left column) and $\beta=0.9$ (the right column).
        The corrugation depth is $d_3=0.5$ cm (the top row), $d_3=1$ cm (the middle row) and $d_3=1.5$ cm (the bottom row).
        Other parameters: $k_0=1$ cm$^{-1}$, $d=0.05$ cm, \linebreak $d_1=0.01$ cm.}
	\end{figure*}

\begin{figure}[h]
		\centering
		\includegraphics[]{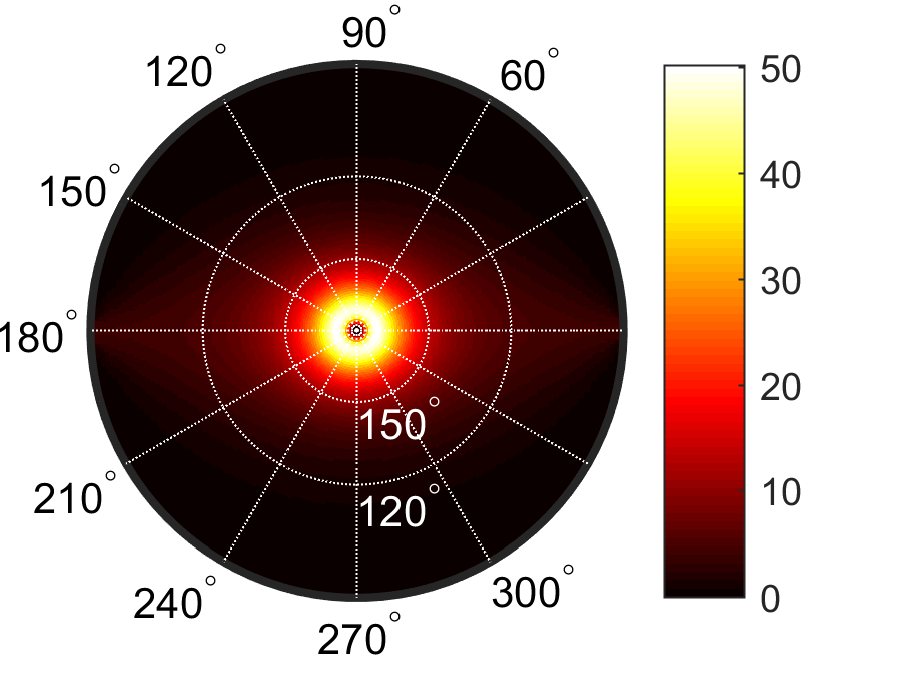}
		\caption{\label{Fig:4} The same as in Fig.~%
		\ref{Fig:3} for $\beta=0.99$ and $d_3=1.5$.}
	\end{figure}

	Fig.~%
	\ref{Fig:3} illustrates the dependences of spectral angular density $W_{\omega\Omega}^{(v)}$ (in $q^2|\tilde{\kappa}|^2/c$ units) on the angles of spherical coordinate system~%
	\eqref{eq:4.2} for the different values of bunch velocity $\beta$ and corrugation depth $d_3$.
	Angle $\theta$ varies along the radius of the circle (from $\theta=90^{\circ}$ at the edge to $\theta=180^{\circ}$ at the center), $\varphi$ is the polar angle.
	As follows from the plots, a maximum of the energy density is observed in the plane $y=0$ that corresponds to the polar angles $\varphi=0^{\circ}$ and $\varphi=180^{\circ}$ (these are the directions along the corrugation grooves, see Fig.~%
	\ref{Fig:1}).
	One can see that the deeper the corrugated structure is, the faster the energy density decreases with a distance from the plane $y=0$.
	This can be explained by the fact that impedance $\eta^m$ increases with an increase in the depth (see Eqs.~%
	\eqref{eq:2.3},~%
	\eqref{eq:2.4}), which results in a decrease in the conductivity in the direction perpendicular to the corrugation grooves ($y$-direction), whereas the conductivity along the grooves ($x$-direction) is perfect under the approximation considered.
	The maximum value of the energy density increases with increasing in the bunch velocity and does not depend on the corrugation depth.
	In the half-plane $\varphi=const$, the value of $\theta$ corresponding to the maximum energy density increases with an increase in the bunch velocity.
	In particular, for $\beta=0.7$ (the plots in the left column) the energy density has a maximum in a direction close to $\theta=90^{\circ}$, i.e. the radiation propagates in the vicinity of the structure plane (this is true for bunches with $\beta<0.7$ as well).
	For $\beta=0.9$ (the plots in the right column) the direction of the radiation maximum is $\theta=151^{\circ}$.
	In the limiting case, when $\beta\rightarrow1$, the energy density has a maximum in the direction opposite to the trajectory of the bunch motion ($\theta\rightarrow180^{\circ}$). 
	In this situation, the distribution of $W_{\omega\Omega}^{(v)}$ does not practically depend on the corrugation depth, which results in a symmetry with respect to angle $\varphi$ (in a neighborhood of the radiation maximum).
	An example for $\beta=0.99$ is given in Fig.~%
	\ref{Fig:4}.

	\section{Surface waves}
	\label{Sec5}
	
	In this section, we focus on analyzing the surface waves.
	As was mentioned, the surface waves are the contributions of poles to Fourier-integrals~%
	\eqref{eq:3.8}.
	The poles are the ones of coefficients $R_x$ and $R_z$, they can be found from the dispersion equation (see Eqs.~%
	\eqref{eq:3.9},~%
	\eqref{eq:3.10})
	\begin{align}
		\label{eq:5.1}
		\tag{5.1}
		k_0k_{z0}-i\left(k_0^2-k_x^2\right)\eta_0^m=0.
	\end{align}
	The solutions of~%
	\eqref{eq:5.1} are
	\begin{align}
		\label{eq:5.2}
		\tag{5.2}
		&k_x=\pm k_{xp}^{(1)}=\pm k_0\sqrt{1-g^2},
	\end{align}
	where
	\begin{align}
		\label{eq:5.3}
		\tag{5.3}
		g^2=\frac{\operatorname{sgn}\left(\eta_0^m\right)}{2(\eta_0^{m})^2}\left(\sqrt{1+\frac{4(\eta_0^{m})^2k_y^2}{k_0^2}}-\operatorname{sgn}\left(\eta_0^m\right)\right).
	\end{align}
	It is important to note that the contributions of~%
	\eqref{eq:5.2} are surface waves if the poles are real.
	As is clear from~%
	\eqref{eq:5.2}, the realness of the poles is determined by the condition $g^2<1$ which can be rewritten in the form
    \begin{align}
    	\label{eq:5.4}
    	\tag{5.4}
    	k_y^2<k_0^2\left[1+(\eta_0^{m})^2\right].
    \end{align}

    We note that coefficients $R_x$ and $R_z$ have also other peculiarities.
    These are the poles $k_x{=}{\pm}k_{xp}^{(2)}{=}{\pm}i\sqrt{k_y^2{+}k_0^2\left(1{-}\beta^2\right)\beta^{-2}}$ which are the solutions of the equation $k_0^2{-}\beta^2k_{z0}^2{=}0$.
    However, poles $\pm k_{xp}^{(2)}$ are purely imaginary for any real values of $k_y$ and $k_0$.
    This means that their contributions to~%
    \eqref{eq:3.8} can be neglected under the condition $k_0\beta^{-1}\sqrt{1{-}\beta^2}|x|{\gg}1$.
    Further, we will assume that $|x|$ is sufficiently great to satisfy this inequality.
    
    The contributions of poles $k_{xp}^{(1)}$ give the following expression for the Fourier-transform of the Hertz potential of the surface wave (these contributions will be marked with index ``s''):
    \begin{align}
    	\label{eq:5.5}
    	&\left\{\Pi_{x\;\omega}^{(s)} \atop \Pi_{z\;\omega}^{(s)}\right\}=2\pi i\operatorname{sgn}\left(x\right)\underset{k_x=\pm k_{xp}^{(1)}}{\operatorname{Res}}\left\{\Pi_{\omega x}^{(r)} \atop \Pi_{\omega z}^{(r)}\right\} \nonumber \\
    	&\!=\frac{q\tilde\kappa\beta}{c}\!\!\int\limits_{-\infty}^{+\infty}\!dk_y\!\left\{\begin{aligned}
    	\eta_0^mk_0^{-1}\left(ik_0-\beta\sqrt{k_y^2-k_0^2g^2}\right)\;\; \\
    	{-}\operatorname{sgn}\left(x\right)\left(1{+}i\beta g^2\eta_0^m\right)\!\left(1{-}g^2\right)^{-\frac{1}{2}}\end{aligned}\right\}\nonumber \\
    	\tag{5.5}
    	&\times\frac{\operatorname{exp}\left(ik_0\sqrt{1-g^2}|x|+ik_yy-\sqrt{k_y^2-k_0^2g^2}|z|\right)}{k_0^2+\beta\left(k_y^2-k_0^2g^2\right)}.
    \end{align}
    Using relations~%
    \eqref{eq:3.3} we obtain the Fourier-transforms of the field components:
    \begin{align}
    	\label{eq:5.6}
    	\tag{5.6}
    	&\left\{\begin{aligned}
    		E_{x\;\omega}^{(s)} \\
    		E_{y\;\omega}^{(s)} \\
    		E_{z\;\omega}^{(s)}\end{aligned}\right\}=\!\!\int\limits_{-\infty}^{+\infty}\!dk_y\left\{\begin{aligned}
    		E_{x\;\omega,k_y}^{(s)} \\
    		E_{y\;\omega,k_y}^{(s)} \\
    		E_{z\;\omega,k_y}^{(s)}\end{aligned}\right\}\operatorname{exp}\left(ik_yy\right),
    \end{align}
    \begin{align}
    	\label{eq:5.7}
    	\tag{5.7}
    	&\left\{\begin{aligned}
    		H_{x\;\omega}^{(s)} \\
    		H_{y\;\omega}^{(s)} \\
    		H_{z\;\omega}^{(s)}\end{aligned}\right\}=\!\!\int\limits_{-\infty}^{+\infty}\!dk_y\left\{\begin{aligned}
    		H_{x\;\omega,k_y}^{(s)} \\
    		H_{y\;\omega,k_y}^{(s)} \\
    		H_{z\;\omega,k_y}^{(s)}\end{aligned}\right\}\operatorname{exp}\left(ik_yy\right),
    \end{align}
    where
    \begin{align}
    	\label{eq:5.8}
    	\notag
    	&\left\{\begin{aligned}
    		E_{x\;\omega,k_y}^{(s)} \\
    		E_{y\;\omega,k_y}^{(s)} \\
    		E_{z\;\omega,k_y}^{(s)}\end{aligned}\right\}
    	=\frac{q\tilde\kappa\beta}{c}\frac{\operatorname{exp}\left(ik_0\sqrt{1-g^2}|x|-\sqrt{k_y^2-k_0^2g^2}|z|\right)}{\sqrt{1-g^2}\left[k_0^2+\beta\left(k_y^2-k_0^2g^2\right)\right]} \\
    	\tag{5.8}
    	&\!\times\!\left\{\begin{aligned}
    		&\qquad\qquad\quad\: ik_0\sqrt{1-g^2}\left(k_0g^2\eta_0^m-\sqrt{k_y^2-k_0^2g^2}\right) \\
    		&\;\,\quad
    		{-}i\operatorname{sgn}\left(x\right)k_y\left[k_0\eta_0^m\left(1-g^2\right)+\left(1+i\beta\eta_0^m\right)\sqrt{k_y^2-k_0^2g^2}\right] \\
    		&\!
    		{-}\operatorname{sgn}\left(x\right)\!\left[k_0^2{+}i\beta\eta_0^mk_y^2{+}\!\sqrt{k_y^2{-}k_0^2g^2}\!\left(\sqrt{k_y^2{-}k_0^2g^2}{+}k_0\eta_0^m\left(1{-}g^2\right)\right)\right]\end{aligned}\right\}\!, \\
    	&\notag \\
    	\notag
    	&\left\{\begin{aligned}
    		H_{x\;\omega,k_y}^{(s)} \\
    		H_{y\;\omega,k_y}^{(s)} \\
    		H_{z\;\omega,k_y}^{(s)}\end{aligned}\right\}
    	=\frac{q\tilde\kappa\beta}{c}\frac{\operatorname{exp}\left(ik_0\sqrt{1-g^2}|x|-\sqrt{k_y^2-k_0^2g^2}|z|\right)}{k_0^2+\beta\left(k_y^2-k_0^2g^2\right)} \\
    	\label{eq:5.9}
    	\tag{5.9}
    	&\!\times\!\left\{\begin{aligned}
    		&-\operatorname{sgn}\left(x\right)k_0k_y\left(1+i\beta\eta_0^mg^2\right)\left(1-g^2\right)^{-\frac{1}{2}} \\
    		&\quad\;\:
    		k_0^2+\eta_0^m\left(k_0\sqrt{k_y^2-k_0^2g^2}+i\beta k_y^2\right) \\
    		&\quad\;\:
    		-k_y\eta_0^m\left(ik_0-\beta\sqrt{k_y^2-k_0^2g^2}\right)\end{aligned}\right\}.
    \end{align}

    Further, we analyze the energy losses of the bunch by the excitation of the surface waves.
    The energy flow of the surface wave through a unit area of a half-plane $x=const$ ($z<0$) over the entire time can be calculated as follows:
    \begin{equation}
    	\label{eq:5.10}
    	\tag{5.10}
    	\frac{d^2W^{(s)}}{dydz}\equiv W_x^{(s)}=\int\limits_{-\infty}^{+\infty}dtS_x.
    \end{equation}
    Here, $S_x=\left(c/4\pi\right)\left(E_y^{(s)}H_z^{(s)}-E_z^{(s)}H_y^{(s)}\right)$ is the $x$-component of Poynting vector.
    Then, we write the components of the surface wave as Fourier-integrals $F\left(\vec{r},t\right)=\iint_{-\infty}^{+\infty}d\omega dk_y F_{\omega,k_y} \exp\left(ik_yy-i\omega t\right)$ and substitute them into~%
    \eqref{eq:5.10}.
    Making some analytical calculations
    we derive from~%
    \eqref{eq:5.10} the equation
    \begin{equation}
    	\label{eq:5.11}
    	\tag{5.11}
    	W_x^{(s)}=\int\limits_{0}^{+\infty}W_{x\;\omega}^{(s)}\;d\omega,
    \end{equation}
    where
    \begin{align}
    	\label{eq:5.12}
    	\tag{5.12}
    	W_{x\;\omega}^{(s)}{=}{-}c\!\!\iint\limits_{-\infty\;\;\;}^{\;\;\;\;+\infty}\!\!dk_ydk'_y\operatorname{cos}\left[\left(k_y{+}k'_y\right)\!y\right]\!\operatorname{Re}\left(E_{y\;\omega,k_y}^{(s)}H_{z\;\omega,k'_y}^{(s)*}{+}E_{z\;\omega,k_y}^{(s)}H_{y\;\omega,k'_y}^{(s)*}\right)
    \end{align}
    is a spectral density of the energy passing through a unit area in $x$-direction (along the corrugation grooves).
    In~%
    \eqref{eq:5.12}, the Fourier-transforms of the field components are defined by Eqs.~%
    \eqref{eq:5.8},~%
    \eqref{eq:5.9}.
    
    By analogy, one can calculate the energy flow through a unit area of a half-plane $y=const$ over the entire time.
    In this case, the energy flow is determined by the equation
    \begin{equation}
    	\label{eq:5.13}
    	\tag{5.13}
    	\frac{d^2W^{(s)}}{dxdz}\equiv W_{y}^{(s)}=\int\limits_{-\infty}^{+\infty}dtS_y,
    \end{equation}
    where $S_y=\left(c/4\pi\right)\left(E_z^{(s)}H_x^{(s)}-E_x^{(s)}H_z^{(s)}\right)$.
    As a result, we have the following expression for the spectral density of the energy passing through a unit area in $y$-direction (perpendicularly to the corrugation grooves):
    \begin{align}
    	\label{eq:5.14}
    	\tag{5.14}
    	W_{y\;\omega}^{(s)}{=}{-}c\!\!\iint\limits_{-\infty\;\;\;}^{\;\;\;\;+\infty}\!\!dk_ydk'_y\operatorname{sin}\left[\left(k_y{+}k'_y\right)\!y\right]\!\operatorname{Im}\left(E_{x\;\omega,k_y}^{(s)}H_{z\;\omega,k'_y}^{(s)*}{-}E_{z\;\omega,k_y}^{(s)}H_{x\;\omega,k'_y}^{(s)*}\right).
    \end{align}
    \begin{figure*}[!h]
    	\vspace{-70pt}
    	\begin{center}
    		\begin{minipage}{0.51\linewidth}
    			\includegraphics[]{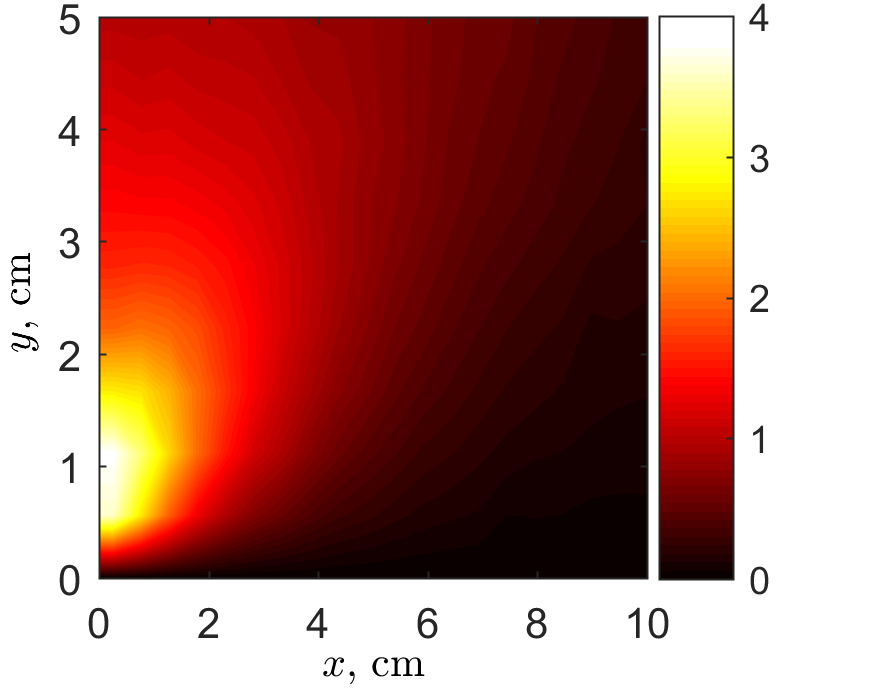}
    		\end{minipage}
    		\begin{minipage}{0.48\linewidth}
    			\includegraphics[]{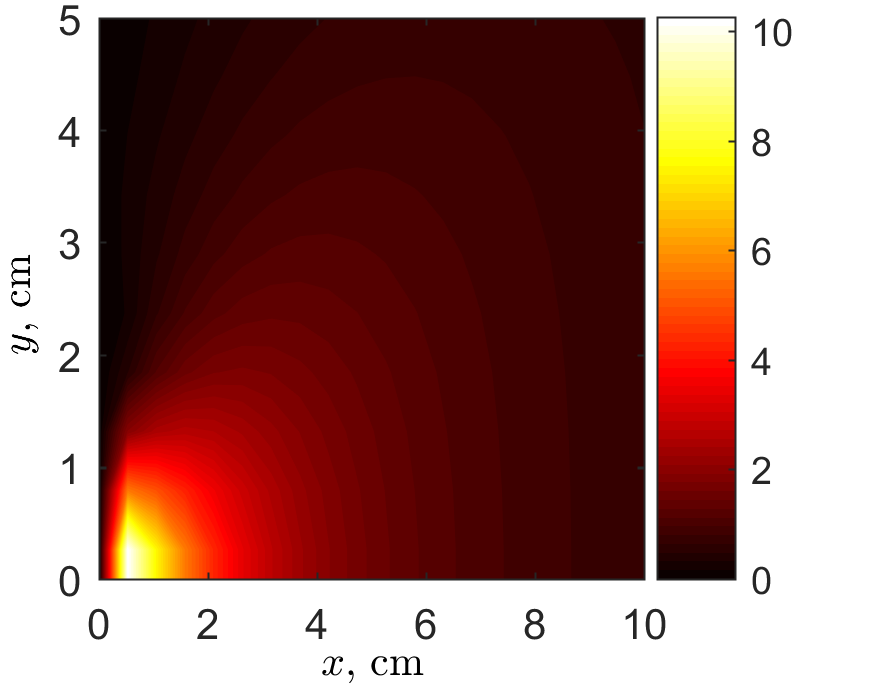}
    		\end{minipage}
    		\begin{minipage}{0.51\linewidth}
    			\includegraphics[]{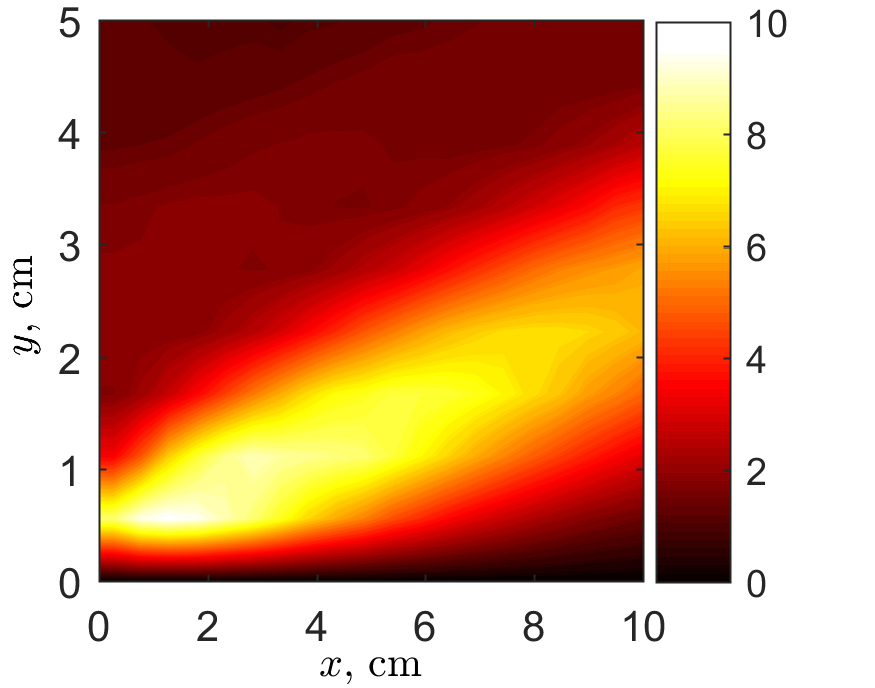}
    		\end{minipage}
    		\begin{minipage}{0.48\linewidth}
    			\includegraphics[]{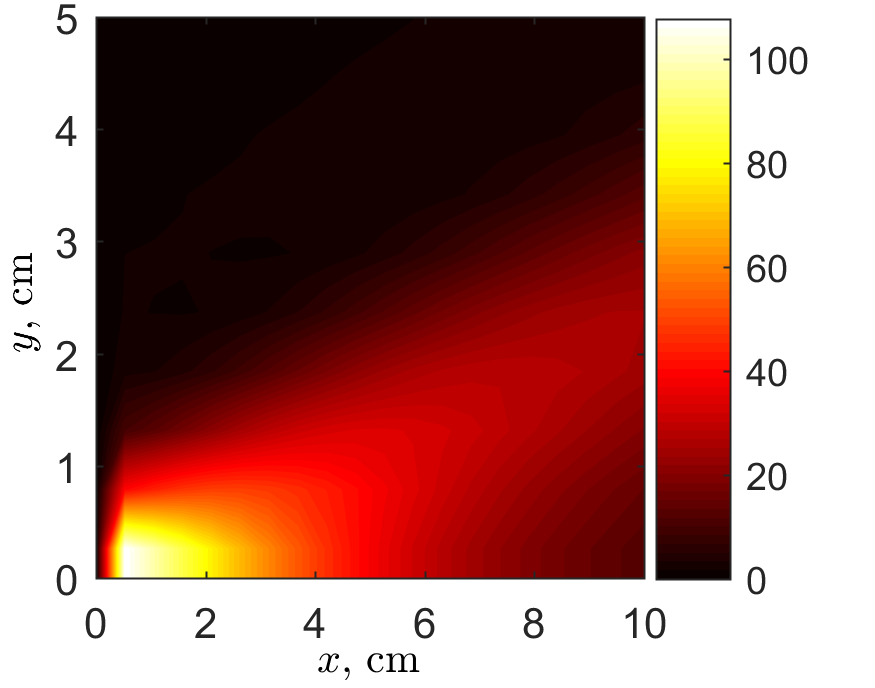}
    		\end{minipage}
    		\begin{minipage}{0.51\linewidth}
    			\includegraphics[]{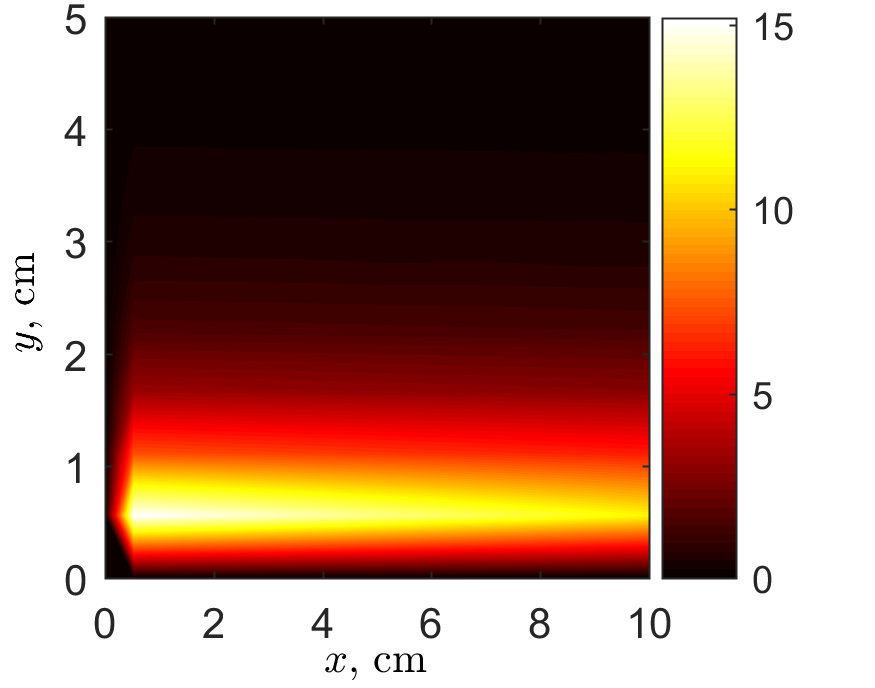}
    		\end{minipage}
    		\begin{minipage}{0.48\linewidth}
    			\includegraphics[]{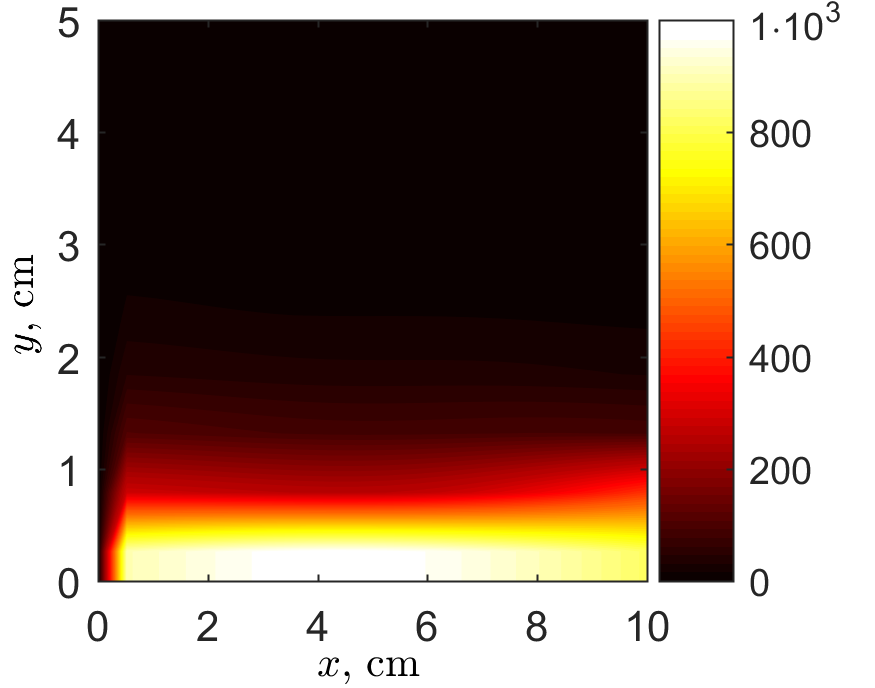}
    		\end{minipage}
    	\end{center}
    	\vspace{-10pt}
    	\caption{\label{Fig:5} Spectral densities $W_{x\;\omega}^{(s)}$ (the right column) and $W_{y\;\omega}^{(s)}$ (the left column) depending on coordinates $x$ and $y$ in the plane $z=-0.5$ cm.
    		The corrugation depth is $d_3=0.5$ cm (the top row), $d_3=1.3$ cm (the middle row) and $d_3=1.5$ cm (the bottom row).
    		Other parameters: $\beta=0.99$, $k_0=1$ cm$^{-1}$, $d=0.05$ cm, $d_1=0.01$ cm.
    		The spectral density is given in $q^2|\tilde{\kappa}|^2/(c \,\cdot\,$cm$^2)$ units.}
    \end{figure*}
    Below, we demonstrate the results of numerical calculations of Eqs.~%
    \eqref{eq:5.12} and ~%
    \eqref{eq:5.14}.
    Fig.~%
    \ref{Fig:5} illustrates the dependences of spectral densities $W_{x\;\omega}^{(s)}$ (the right column) and $W_{y\;\omega}^{(s)}$ (the left column) on coordinates $x$ and $y$ in the plane $z=-0.5$ cm for the different values of corrugation depth $d_3$.
    According to the plots, the distribution of the energy density is highly sensitive to the value of the depth.
    The concentration of the energy along the corrugation grooves ($x$-direction) increases with increasing in the depth.
    The explanation of this fact is the same as in the previous section, where the volume radiation was analyzed: the conductivity of the structure in $y$-direction decreases with increasing in the depth, whereas the conductivity in $x$-direction is perfect under the approximation considered (see Eqs.~%
    \eqref{eq:2.3},~%
    \eqref{eq:2.4}).
    Therefore, relation $W_{x\;\omega}^{(s)}/W_{y\;\omega}^{(s)}$ increases as well, and in the case of a relatively small conductivity in the direction perpendicular to the grooves we have \linebreak $W_{x\;\omega}^{(s)}\gg W_{y\;\omega}^{(s)}$ (the plots in the bottom row of Fig.~%
    \ref{Fig:5}).
    It should be also noted that the surface radiation is much more powerful than the volume one when the conductivity in $y$-direction is relatively small.
    In particular, this can be observed by comparing the plots in Fig.~%
    \ref{Fig:4} and in the right column of Fig.~%
    \ref{Fig:5} (we note again that the maximum value of the energy density for the volume radiation does not depend on the depth for the constant value of $\beta$).

    \section{Conclusion}
    \label{Sec6}
    
    We investigated the electromagnetic radiation from the thin bunch passing through the corrugated planar structure.
    The radiation was analyzed in the ``longwave'' range, i.e. it was assumed that the wavelengths under consideration are much greater than the structure period.
    This assumption allowed using the EBC method to obtain the general solution of the problem.
    We assumed as well that the structure is deeply corrugated, that is, the depth is of the same order as an inverse wavenumber.
    The asymptotic investigation of the general solution showed that the bunch generates volume radiation and surface waves propagating along the structure plane.
    We obtained the field components for both types of the radiations and presented them in the form of Fourier-integrals.
    Then, we considered the energy losses of the bunch and derived expressions for the spectral densities of the energy for both volume radiation and surface waves.
    Finally, we presented and analyzed the dependences of the energy densities on the bunch velocity and the corrugation parameters.

\section{Acknowledgments}

This work was supported by the Russian Science Foundation (Grant No.~18-72-10137).

	\bibliographystyle{model1a-num-names}
	\bibliography{Manuscript}

\end{document}